\documentclass[11.5pt,a4paper,twoside]{article}

\usepackage[utf8]{inputenc}
\usepackage[english]{babel}
\usepackage{geometry}
\usepackage{float}
\usepackage{amsmath, amsthm, amssymb, amsfonts}
\usepackage{bm, bbm}
\usepackage[colorlinks=true,linkcolor=blue, linktoc=page, urlcolor=blue,citecolor=magenta]{hyperref}
\usepackage{xcolor}
\usepackage{graphicx}
\usepackage{comment}
\usepackage{cite}
\usepackage{caption}
\usepackage{subcaption}
\usepackage{ifdraft}
\usepackage[section]{placeins}

\numberwithin{equation}{section}

\usepackage[normalem]{ulem}

\begin{document}

\allowdisplaybreaks

\begin{titlepage}
    %
    \begin{flushright}\vspace{-2cm}
        {\small \today }
    \end{flushright}
    \bigskip

    \begin{center}
        %
        {\LARGE \textbf{Event Horizon of a Charged Black Hole Binary Merger }}\\
        \line(2,0){450}
        \bigskip
        {\textbf{
            D.~Mar\'in Pina$^\dagger$\footnote{danielmarin@icc.ub.edu},
            M.~Orselli$^{\ddagger\,*}$\footnote{orselli@nbi.dk},
            D.~Pica$^{\ddagger\,*}$\footnote{daniele.pica@studenti.unipg.it}}}\vspace{2pt}
            
        \bigskip
        \medskip
        
        \textit{{}$^\dagger$ Departament de F\'isica Qu\`antica i Astrof\'isica, Institut de Ci\`encies del Cosmos, Universitat de Barcelona, Mart\'i i Franqu\`es 1, E-08028 Barcelona, Spain}\\
        \medskip
        \textit{{}$^\ddagger$ Niels Bohr Institute, Copenhagen University,\\ Blegdamsvej 17, DK-2100 Copenhagen \O{}, Denmark}\\ 
        \medskip
        \textit{{}$^*$ Dipartimento di Fisica e Geologia, Universit\`a di Perugia, I.N.F.N. Sezione di Perugia, \\ Via Pascoli, I-06123 Perugia, Italy }
        \line(2,0){450}
        \vspace{10mm}
    

\begin{abstract}
\noindent 
We analyse the formation and evolution of the event horizon of a black hole binary merger when the black holes are charged. We find that the presence of charge influences the properties of the merger and can be useful for investigating the validity of various theories of modified gravity and several proposals for dark matter candidates. It can moreover give insights into various aspects of astrophysical phenomena involving black holes, such as degeneracies in the gravitational wave parameter determination. We perform our analysis both analytically and numerically, in $D=4$ dimensions, in the Extreme Mass Ratio (EMR) limit and compare the results. The development of analytical results in the EMR limit is of uttermost importance in view of the upcoming observations of the LISA interferometer. We then use our analysis to describe how the horizon evolves in time during the merger and to investigate the growth in the area of the event horizon and the duration of the merger. We moreover provide a numerical solution valid in arbitrary dimensions $D\geq 4$ which could be of interest in the context of the AdS/CFT correspondence or for examining possible extensions of general relativity.

\end{abstract}
    \end{center}
    \setcounter{footnote}{0}
\end{titlepage}

\tableofcontents
\vspace{0.5 cm}


\section{Introduction}
\label{Sec:Intro}

One of the greatest achievements in physics of recent years is the detection, by the LIGO-Virgo-KAGRA collaboration, of gravitational wave signals emitted by binary systems of two black holes \cite{LIGOScientific:2016aoc}, two neutron stars \cite{LIGOScientific:2017vwq} and a black hole and a neutron star \cite{LIGOScientific:2021qlt}.
The study of the physics and of the dynamics of binary systems has thus become of paramount importance. In this context, one of the properties which is interesting to study and from which one could extract relevant information for gravitational wave astrophysics is the formation and evolution of the event horizon in the merger of a black hole binary system. A step forward in this direction has been made in \cite{Emparan:2016ylg} for the case of neutral spherically symmetric black holes, subsequently generalized in \cite{Emparan:2017vyp} to rotating neutral black holes. Moreover, the event horizon formation and properties during the merger of a neutron star-black hole binary system have been studied in~\cite{Emparan:2020uvt}.

Recently (see for example \cite{Maldacena:2018gjk,zajaček2019electric, Maldacena:2020sxe, Nomura:2021efi, Dias:2021yju, Carullo:2021oxn, Komissarov:2021vks, Pereniguez:2021xcj}) there has been some interest in including the presence of charge in the study of various properties of black holes and, in particular, of black hole binary systems. This could in fact be relevant not only for constructing waveform models that could be used to describe gravitational wave signals in sensible and realistic astrophysical situations \cite{Dias:2021yju, Carullo:2021oxn}, but it could also provide essential information for investigating the validity of certain theories of modified gravity \cite{Bozzola:2020mjx} and various dark matter scenarios \cite{Cardoso:2016olt}.
This is in contrast with the fact that, even though they might posses a small amount of electric charge, astrophysical black holes are typically considered neutral (See e.g.~\cite{Wald:1974np, Gibbons:1975kk, Blandford:1977ds}) and therefore all information one extrapolates from analysing physical phenomena, such as gravitational radiation emitted during the merger of a binary black hole system, is obtained using the Kerr metric. Nonetheless, as already mentioned, allowing for the presence of this extra parameter could be relevant in providing further tests on how the presence of (a small amount of) electric charge would affect the processes involving black holes, as well as being important for investigating the validity of various exotic models of minicharged dark matter~\cite{Cardoso:2016olt}, where the charge is due to primordial magnetic monopoles~\cite{Preskill:1984gd, Bozzola:2020mjx}, and other possible scenarios.
Therefore, in this paper we use the term charge to refer to electric or magnetic charge \cite{Maldacena:2020skw,1978ApJ...220..743B}, something more exotic such as minicharged dark matter \cite{DERUJULA1990173, Cardoso_2016} or a new scalar or vector field in the theory of gravity \cite{2006, Maselli:2021men}.

With this motivation in mind, we study the merger of a binary system of two black holes where at least one of them is a charged black hole described by the Reissner-Nordstr\"om (RN) solution of vacuum Einstein's equations  \cite{Einstein:1916vd}. Our aim is to investigate the formation and evolution of the event horizon of such a binary system, and we will perform our analysis both analytically and numerically. 

Obtaining an analytical description of the formation of the event horizon in a binary system merger is typically a very challenging task. However, as shown in \cite{Emparan:2016ylg}, things simplify considerably in the so-called Extreme-Mass-Ratio (EMR) limit where one of the two black holes is much smaller than the other.
If $m$ is the mass of the small black hole and $M$ is the one of the large black hole (in units where $G=c=\epsilon_0=\mu_0=1$), this limit consists in sending the ratio $m/M\rightarrow 0$. This is observationally tied to the merger of supermassive black holes (SMBH) and stellar-mass black holes, and detecting them is one of the core goals of the upcoming LISA mission \cite{Babak:2017tow}.

The EMR limit is usually realised in gravitational wave astrophysics by sending the mass $m$ of the small black hole to zero while keeping the mass $M$ of the large black hole fixed. This approach is then used to extract the gravitational wave signal emitted during the merger of the binary system. In this realisation of the EMR limit, the small black hole can be regarded as a test particle, and one can then calculate the gravitational wave emitted by the system using e.g. a multipolar Regge-Wheeler-Zerilli perturbative approach \cite{Bernuzzi:2010ty,Nagar:2006xv} or an Effective One Body  approach \cite{Buonanno:1998gg, Damour:2008te, Albanesi:2021rby, Nagar:2021gss, Bernuzzi:2010xj, Damour:2009zoi}. However, as pointed out in~\cite{Emparan:2016ylg}, sending $m$ to zero while keeping $M$ fixed is not suitable if one wants to study the evolution in time of the event horizon of a binary black hole merger. In fact, by sending the mass of the small black hole to zero, and thus treating it as a point-like object, one loses all information about the physics at the scale $m$, making it not appropriate for studying the evolution of the event horizon during the merger. When the small black hole is treated as a point-like object we are in fact not able to study its geometrical structure. This means that we are no longer able to find the null hypersurface that identifies the event horizon.

To overcome this problem, we have to find an alternative way of taking the EMR limit such that we preserve the information about the event horizon of the small black hole. To do this, it is instead convenient to send $M\rightarrow \infty$ while keeping $m$ fixed \cite{Emparan:2016ylg}. This has the advantage that, in the rest frame of the small black hole, we can neglect the curvature of spacetime over distances $\ll M$. This means that, when the small black hole is very close to the large black hole, namely it is at a distance 
from the event horizon of the large black hole much smaller than the size $M$ of the large black hole, the curvature of the large black hole becomes negligible but, at the same time, it is still possible to identify its event horizon, which becomes an infinite acceleration horizon (i.e. a Rindler horizon)\cite{Emparan:2016ylg}. 

In this paper, the small black hole is a charged, non-spinning black hole, so the space-time around it is well described by the RN solution. The presence of the charge introduces an extra parameter, compared to the case of the merger between two neutral black holes, which influences the properties of the merger and which can be used to investigate in an alternative way the event horizon formation and evolution. 

The procedure we will use to find the event horizon in the early time of the merger of a charged black hole binary system in the EMR limit 
is quite simple.
We start by considering the situation just before the merger, from the rest frame of the small black hole, where the small black hole is very close (at distances $\ll M$) to the large one. In this way one can ignore the curvature of the large black hole and one can describe the spacetime around the small black hole using the RN solution (see Eq.~\eqref{RN}). Using that the event horizon of the large black hole in the limit $m/M\rightarrow 0$ becomes an infinite acceleration horizon, we  construct the event horizon of the binary system starting from the configuration of the event horizon on $\mathcal{I}^+$, which is a congruence of light rays that forms a planar surface at asymptotic null infinity, and then following back in time a congruence of null geodesics that, starting from the small, charged black hole, reaches a planar horizon at large distance \cite{Emparan:2016ylg}. See Fig.~\ref{plot:rn:sliceplot} for a visualisation of this process.

This approach allows us to extract the most relevant features of the event horizon of
a black hole binary merger, such as the duration of the merger, the growth in the area of the small black hole and the presence of a line of caustics.\footnote{A line of caustics is always present in the event horizon of a binary black hole merger. It is due to the strong gravitational field that causes light rays to intersect one another in caustic points, which form a line of caustics. These points are the first to enter the event horizon in the early time of the merger.}

The paper is organized as follows. In Sec.~\ref{Sec:RN} we study analytically the evolution of the event horizon of a black hole binary system in $D=4$ dimensions in the EMR limit, when the smaller object is a charged black hole. We describe the procedure used in order to extract some of the most interesting features of the evolution of the event horizon, such as the presence of a line of caustics, the growth of the area of the event horizon and the duration of the merger. We moreover argue that our results are valid for the merger of a small charged black hole with any type of large black hole, in the EMR limit, independently of the spin or charge of the large black hole.
In Sec.~\ref{Sec:Num}, we study the same problem using a different procedure, which uses the Hamiltonian form of the geodesics equations and which is particularly useful for numerically solving the differential equations at hand. This procedure has also the advantage that can be easily generalized to arbitrary dimensions $D > 4$. We do this in  Sec.~\ref{Sec:Dim} where we apply it to the specific case of $D =5$.  

We moreover show the similarities and the main differences with the computation in $D=4$ dimensions. Sec.~\ref{Sec:conclusions} contains a summary of our results and some concluding remarks.
Throughout the paper we use units where $G=c=\epsilon_0=\mu_0=1$, unless otherwise specified.

\section{Analytical solution in $D=4$}
\label{Sec:RN}
In this section we study analytically the merger of two black holes when at least one of them is charged, in the EMR limit. This analysis allows us to describe the evolution in time of the event horizon of the system
and to study the growth in the area of the event horizon and the duration of the merger. 

Charged black holes are described by the generalization of the RN solution which, in $D=4$ dimensions and using Schwarszschild quasi-spherical coordinates, can be written as
\begin{equation}
\label{RN}
ds^2=-\Delta(r) dt^2+\Delta(r)^{-1} dr^2+r^2 d\Omega^2,
\end{equation}
with 
\begin{equation}
\label{delta}
\Delta(r)=1-\frac{2m}{r}+\frac{Q^2}{r^2}.
\end{equation}
Here, $Q$ denotes the electric or magnetic charge, a combination of the two or, in general, any type of parameter with the same coupling of the charge in the Einstein-Maxwell theory. Moreover, $d\Omega^2=d\theta^2+\sin^2\theta d\phi^2$ and $m$ is the mass of the black hole. 

The location of the event horizon is obtained by solving $\Delta(r)=0$, which yields $r_{+}=m + \sqrt{m^2-Q^2}$.~\footnote{$\Delta(r)=0$ has two solutions given by $r_{\pm}=m \pm \sqrt{m^2-Q^2}$ which represent the outer and inner horizon, respectively. Only the former is an event horizon, with $r_-$ being a Cauchy horizon.}
Here we are only interested in examining situations where $|Q|\leq m$, as only in this case there is an event horizon. The limiting case $|Q|=m$ identifies an \textit{extremal} Reissner-Nordstr\"om black hole, where the event horizon is located at $r_{\rm{EH}}=m$. This case represents a black hole with the maximum amount of charge, equal to it mass. Since the mass of a black hole gives also its dimension, this means that an extremal black hole is the smallest possible configuration. This type of black holes correspond to an unstable (and not physical) situation and, as such, they are not found in Nature. Nonetheless, extremal black holes are very interesting from a purely theoretical and analytical point of view. They typically play a crucial role in the context of supersymmetric theories where they often represent stable solutions and are used as a valuable tool to investigate important problems such as, for example, obtaining a microscopic description of the Bekenstein Hawking entropy formula for black holes in terms of D-branes configurations~\cite{Strominger:1996sh}. Given their interest from a theoretical perspective, in this paper we will also consider the extremal case $|Q|= m$.

\subsection{Derivation of the characteristic equations: Case $|Q|< m$}
\label{regular}
We start by considering a regular RN black hole with $|Q|< m$. The extremal case $|Q|= m$ can be easily obtained by taking the limit $|Q|\to m$ in the results obtained here and it will be discussed separately in Sec.~\ref{Sec:extr}.

We want to study the merger of a charged binary black hole system in the EMR limit, which we take by sending the mass $M$ of the large black hole to infinity while keeping the mass $m$ of the small black hole
fixed. As already explained, this is the correct way of employing the EMR limit in order to being able to retain information about the event horizon of the small black hole~\cite{Emparan:2016ylg}.
We investigate this system from the point of view of the rest frame of the small black hole, when the small black hole is a RN black hole and is very close to the large black hole, namely at distances $\ll M$. This provides a set up where the merger of the two black holes can be studied in a simple way, as already explained in~\cite{Emparan:2016ylg} for the neutral case (see also \cite{Emparan:2017vyp} for the case of rotating black holes). In fact, in this description, it can be shown that the symmetries of spacetime are exactly the ones of the RN background, namely spherical symmetry together with a timelike Killing vector $\partial_t$. This is true only in the exact limit $m/M\to 0$. When taking into account corrections of order $m/M$ the symmetries are not exact anymore and the study of the merger becomes much more complicated.
When the small black hole is at distances $\ll M$ from the large black hole and the geometry is given by the metric \eqref{RN}, the event horizon of the large black hole becomes an acceleration horizon with the geometry of a null plane. 
Therefore, to study the merger between the two black holes, we need a null surface that reaches  $\mathcal{I}^+$ with the geometry of an acceleration horizon at a finite retarded time and that represents the configuration of the event horizon of the large black hole in the limit $M\rightarrow \infty$.

This procedure allows us to study the properties of the event horizon of the binary system in the early time of the merger, starting from the situation when the two event horizons are infinitely separated. 

Our aim is to describe how the two event horizons change due to gravitational attraction  from being two separated surfaces until they merge forming a single smooth surface. See Figs.~\ref{plot:rn:sliceplot} and \ref{plot:rnd:sliceplot} for a graphic representation of this process. 

Event horizons are null hypersurfaces, which can be thought of as a congruence of null geodesics, called generators. We know that at future null infinity, $\mathcal{I}^+$, the generators of the event horizon form a planar horizon. The idea is then to trace back these null geodesics from $\mathcal{I}^+$ until they reach a so-called \textit{line of caustics}, which is the set of points where null geodesics focus as they enter the event horizon \cite{Shapiro:1995rr,Lehner:1998gu,Husa:1999nm,Siino}. One can then follow the null geodesics back until they reach these points because, once at the horizon, a null generator can never propagate off it, nor it can ever cross another null generator. In other words, the event horizon is defined by null geodesics passing through each point of the null hypersurface and these generators continue along the horizon forever into the future (in agreement with the cosmic censorship conjecture). The event horizon begins where these generators meet in the past. 

Our goal is now to follow back in time the generators of the event horizon from $\mathcal{I}^+$ until they meet at the caustic points. To do this, we need to consider the geodesic equations followed by a specific set of light-rays in the RN background. In particular since the event horizon at $\mathcal{I}^+$ is a planar horizon, we are looking for a congruence of generators that, in the RN geometry, approach a planar horizon at future null infinity. We have already mentioned that, in the limit $m/M\rightarrow 0$, the spacetime geometry we are considering in the rest frame of the small black hole has the same symmetries as the RN background. Thanks to the spherical symmetry of this background, we can set $\theta=\pi/2$ without loss of generality. The  equations that we want to solve read 
\begin{equation}
\dot{t}=\frac{1}{1-\frac{2m}{r}+\frac{Q^2}{r^2}},
\label{105}
\end{equation}
\begin{equation}
\dot{\phi}=-\frac{q}{r^2},
\label{106}
\end{equation}
\begin{equation}
\dot{r}=\frac{1}{r} \sqrt{r^2-q^2\left(1-\frac{2m}{r}+\frac{Q^2}{r^2}\right)},
\label{107}
\end{equation}
where $q$ represents the ratio between the conserved angular momentum and the energy of the light-ray trajectory and it is called \textit{impact parameter}. Moreover $\dot{t}=\frac{dt}{d\lambda}$, $\dot{\phi}=\frac{d\phi}{d\lambda}$ and $\dot{r}=\frac{dr}{d\lambda}$ with $\lambda$ the affine parameter.

In order to better understand the meaning of the impact parameter $q$, we can define the cartesian-like coordinates
\begin{equation}
x=r\sin(\phi), \hspace{3cm} z=r\cos(\phi).
\label{38}
\end{equation}
In the $(x,z)$ coordinates, asymptotically, all light rays move with $dx=0$,
\begin{equation}
x|_{r \rightarrow \infty}= q + \mathcal{O}\left(r^{-3}\right),
\label{39}
\end{equation}
\begin{equation}
z|_{r \rightarrow \infty}= r +\mathcal{O}\left(r^{-1}\right)
\label{40}
\end{equation}
and the horizon is identified by
\begin{equation}
dt-dz=\mathcal{O}\left(r^{-3}\right).
\label{41}
\end{equation}

\begin{figure}
    \centering
    \includegraphics[width=10cm]{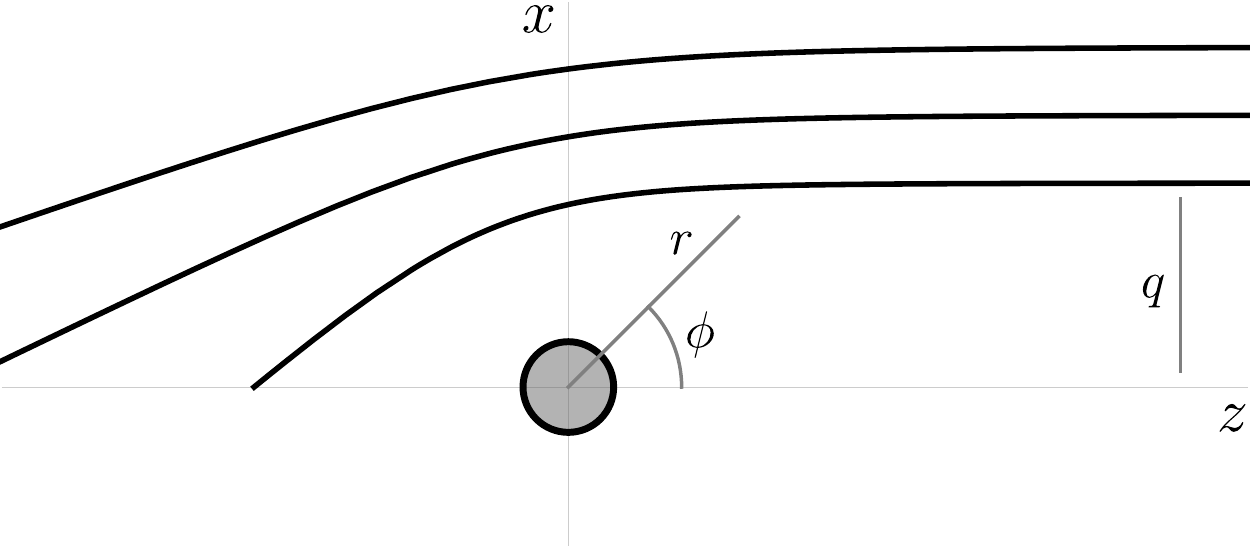}
    \caption{Projection of null generators of the event horizon on the spatial plane $(x,z)$. The black lines are the light rays that move towards $\mathcal{I}^+$. At late times, they move along the $z$ direction as the generators of a Rindler horizon $(dt=dz)$. They are labelled by the impact parameter $q$ at future infinity. This graphic representation is the same of Figure 1 from \cite{Emparan:2016ylg}, which we reproduce here for convenience.} 
    \label{fig:xz}
\end{figure}

As we can see from Fig.~\ref{fig:xz}, the ratio between the angular momentum and the energy of the null geodesics, i.e. the impact parameter $q$, represents the distance at future null infinity between the geodesics and the $z$ axis. The distance between the geodesics and the center of the small black hole, which we consider to be the center of our reference system, is denoted by $r$. Furthermore, studying the geodesics on the plane $\theta=\pi/2$ allows us to choose the $\phi$ axis as the axis along which the collision between the two black holes takes place. To be more exhaustive, $\phi=0$ and $\phi=\pi$ are segments in the plane $\theta=\pi /2$. These two segments define the collision axis along which the collision takes place. Specifically, placing ourselves in the rest frame of the small black hole, the segment $\phi=0$ points in the direction away from the large black hole, while $\phi=\pi$ points in the direction towards the large black hole.

Instead of using $\lambda$, it is convenient to use $r$ as the parameter along the geodesics. With this choice, the equations that we need to solve are 
\begin{equation}
\phi_{q}(r)=\int dr \frac{\dot{\phi}}{\dot{r}}, \hspace{3cm} t_q(r)=\int dr \frac{\dot{t}}{\dot{r}}.
\label{108}
\end{equation}

The next step is to fix the integration constants in Eq.~\eqref{108} by requiring that the null surface which defines the event horizon at $\mathcal{I}^+$ becomes a planar horizon. For $r \rightarrow \infty$, we get 
\begin{equation}
\phi_q(r \rightarrow \infty)=\left.\int dr \frac{\dot{\phi}}{\dot{r}} \vphantom{\big|}\right |_{r \rightarrow \infty}= \alpha_q + \frac{q}{r} + \mathcal{O}(r^{-3}),
\label{109}
\end{equation}
and 
\begin{equation}
t_q(r\rightarrow \infty)=r +2m \log \left(\frac{r}{2m}\right) +\beta_q + \mathcal{O}(r^{-1}).
\label{110}
\end{equation}
We must set $\alpha_q$ and $\beta_q$ to $q$-independent values. This is because we want that the light rays asymptotically move all in the same direction and that they arrive at $\mathcal{I}^+$ at the same retarded time. We choose 
\begin{equation}
\alpha_q=0, \hspace{3cm} \beta	_q=0.
\label{111}
\end{equation}

Evaluating Eq.~\eqref{108} at $q=0$ defines the so-called {\it{central}} geodesics which, going back in time, represents a light ray which starts at $r=r_+$ at $t \rightarrow -\infty$ and moves towards infinity in the direction $\phi=0$.

We find
\begin{equation}
\phi_{q=0}(r)=0,
\label{112}
\end{equation}
\begin{equation}
t_{q=0}(r)=m \log \left(Q^2+r^2- 2mr\right)+\frac{\left(2m^2- Q^2\right) \tan ^{-1}\left(\frac{r-m}{\sqrt{Q^2-m^2}}\right)}{\sqrt{Q^2-m^2}}+r.
\label{113}
\end{equation}

The general, $q$-dependent integrals in Eq.~\eqref{108} have instead the following explicit expressions 
\begin{equation}
\phi_q(r)=-\int \frac{q \hspace{0.2cm}dr}{\sqrt{r^4-q^2r^2+2mq^2r-q^2Q^2}},
\label{114}
\end{equation}
\begin{equation}
t_q(r)=\int \frac{r^4\hspace{0.2cm}dr}{(r^2-2mr+Q^2)\sqrt{r^4-q^2r^2+2mq^2r-q^2Q^2}}.
\label{115}
\end{equation}

We need to solve Eqs.~\eqref{114} and \eqref{115} choosing carefully the values for the integration constants so that the correct asymptotic behavior given in Eq.~\eqref{109} and Eq.~\eqref{110} is reproduced. In order to do so, we begin by rewriting the integrals as 

\begin{equation}
\phi_q(r)=-\int \frac{q \hspace{0.5mm}dr}{\sqrt{(r-x_1)(r-x_2)(r-x_3)(r-x_4)}},
\label{116}
\end{equation}
\begin{equation}
t_q(r)=\int \frac{r^4dr}{(r-r_1)(r-r_2)\sqrt{(r-x_1)(r-x_2)(r-x_3)(r-x_4)}},
\label{117}
\end{equation}
where $x_1, x_2, x_3, x_4$ are the solutions of the quartic equation $r^4-q^2r^2+2mq^2r-q^2Q^2=0$, and $r_{1,2}$ are the solutions of the quadratic equation  $r^2-2mr+Q^2=0$. We have
\begin{eqnarray}
 &&x_{1,2}=\frac{1}{2\sqrt{6}}\left(\xi_1^\frac{1}{2}\mp\xi_2^\frac{1}{2}\right),\cr
 &&x_{3,4}=-\frac{1}{2\sqrt{6}}\left(\xi_1^\frac{1}{2}\pm\xi_3^\frac{1}{2}\right),\cr
&&r_{1,2}=m\mp\sqrt{m^2-Q^2},
\end{eqnarray}
where
\begin{eqnarray}
    &&\xi_1=\frac{2 \sqrt[3]{2} \gamma}{p}+2^{2/3} p+4 q^2,\cr
    &&\xi_2=-\frac{24 \sqrt{6} q^2m}{(\frac{2 \sqrt[3]{2} \gamma}{p}+2^{2/3} p+4 q^2)^{\frac{1}{2}}}-\frac{2 \sqrt[3]{2} \gamma}{p}-2^{2/3} p+8 q^2,\cr
    &&\xi_3=\frac{24 \sqrt{6} q^2m}{(\frac{2 \sqrt[3]{2} \gamma}{p}+2^{2/3} p+4 q^2)^{\frac{1}{2}}}-\frac{2 \sqrt[3]{2} \gamma}{p}-2^{2/3} p+8 q^2,\cr
    &&p=\biggl(\sqrt{q^8 \left(2 q^2+72 Q^2-108m^2\right)^2-4 \left(q^4-12 q^2 Q^2\right)^3}-q^2\left(2 q^2+72 Q^2-108m^2\right)\biggr)^{\frac{1}{3}},\cr
    &&\gamma=q^4-12 q^2 Q^2.
\end{eqnarray}
Since $r_1$ and $r_2$ satisfy $r_1+r_2=2m$, we 
will use this relation to eliminate $r_2$ in favor of $r_1$  when convenient. 
We will use also the relation $x_1+x_2+x_3+x_4=0$ to simplify the expressions in the integrals.

The results of the integrals can then be express in terms of incomplete elliptic integrals of the first, second and third kind, defined as
\begin{equation}
F(x|\bar{m})=\int_0^x \frac{d\theta}{\sqrt{1-\bar{m}\sin^2\theta}},
\label{56}
\end{equation}
\begin{equation}
E(x|\bar{m})=\int_0^x\sqrt{1-\bar{m}\sin^2\theta}d\theta,
\label{57}
\end{equation}
\begin{equation}
\Pi(n;x|\bar{m})=\int_0^x\frac{d\theta}{(1-n\sin^2\theta)\sqrt{1-\bar{m}\sin^2\theta}}.
\label{58}
\end{equation}
When evaluating these expressions one must be careful with the prescription for square root of complex numbers and with the branch cuts in the elliptic functions. Our prescriptions are those implemented in \textit{Mathematica 12}, which we have used for these calculations.

In order to fix the integration constants to the desired values of Eq.~\eqref{111}, it is convenient to use the following relation for elliptic integrals  \cite{10.5555/1098650}
\begin{equation}
\Pi(n;\varphi|\alpha)=-\Pi(N;\varphi|\alpha) + F(\varphi,\alpha)+\frac{1}{2p}\log[(\Delta(\varphi)+p\tan\varphi)(\Delta(\varphi)-p\tan\varphi)^{-1}],
\label{59}
\end{equation}
where 
\begin{equation}
N=n^{-1}\sin^2\alpha, \hspace{5mm} p=[(n-1)(1-n^{-1}\sin^2\alpha)]^{\frac{1}{2}}, \hspace{5mm}  \Delta(\varphi)=(1-\sin^2\alpha\sin\varphi)^{\frac{1}{2}}.
\label{60}
\end{equation}
To use this identity in Eq.~\eqref{58}, we should clarify that $\sin^2\alpha=\bar{m}$ and $\varphi=x$, according to the prescriptions of \textit{Mathematica 12}. After using this identity and fixing the integration constants to the values required in Eq.~\eqref{111}, we get
\begin{equation}
\phi_q(r)=  \frac{2q }{\sqrt{b f}}\biggl(F(y_1|s) -  F(y_2|s)\biggr), 
\label{129}
\end{equation}
and
\begin{eqnarray}
\label{tqr}
t_q(r)&=&\frac{ 1}{(x_2-r_1) (x_2-r_2) \sqrt{bf}}\biggl(F(y_2|s)\biggl(2m a(r_1-x_2) (2r_2+x_1+x_3)
-r_1^2 a f-x_2^2 (x_3x_2+x_1x_4 )\biggr)\cr
&+& x_2  \biggl(2m a(r_1-x_2) -ar_1^2 +x_2^2 (x_1+x_2)\biggr)\biggr)+\sqrt{\frac{(r-x_1)(r-x_3)(r-x_4)}{r-x_2}}- \sqrt{b f} E(y_2|s)\cr
&+&\frac{a}{(m-r_1)\sqrt{b f}}\left[\frac{r_1^4  \Pi \biggl(\frac{(r_1-x_2) c}{(r_1-x_1) f};y_2|s\biggr)}{ (r_1-x_1) (r_1-x_2) }-\frac{ r_2^4 \Pi \biggl(\frac{(x_2-r_2) c}{(x_1-r_2) f};y_2|s\biggr)}{ (r_2-x_1) (r_2-x_2) }\right]\cr
&+& 2m \left[\log \biggl(\frac{\sqrt{(r-x_1) (r-x_2)}+\sqrt{(r-x_3) (r-x_4)}}{\sqrt{(r-x_3) (r-x_4)}-\sqrt{(r-x_1) (r-x_2)}}\biggr)
-2\frac{a  \Pi \biggl(\frac{d}{b};y_2|s\biggr)}{\sqrt{b f}}\right]-c_q,
\end{eqnarray}
where
\begin{eqnarray}
c_q&=&\frac{F(y_1|s)}{(x_2-r_1) (x_2-r_2) \sqrt{bf}}\biggl(2m a(r_1-x_2) (2r_2+x_1+x_3)
-r_1^2 a f-x_2^2 (x_3x_2+x_1x_4 )\biggr)\cr
&+& x_2\left[\frac{  2ma (r_1-x_2)-ar_1^2 +x_2^2 (x_1+x_2)}{(x_2-r_1)  (x_2-r_2) \sqrt{b f}}+1\right]-\sqrt{b f} E(y_1|s)+2m \left[\log \biggl(\frac{2}{x_1+x_2}\biggr)
-2\frac{a  \Pi \biggl(\frac{d}{b};y_1|s\biggr)}{\sqrt{b f}}\right]\cr
&+&\frac{a}{(m-r_1)\sqrt{b f}}\left[\frac{r_1^4  \Pi \biggl(\frac{(r_1-x_2) c}{(r_1-x_1) f};y_1|s\biggr)}{ (r_1-x_1) (r_1-x_2) }-\frac{ r_2^4 \Pi \biggl(\frac{(x_2-r_2) c}{(x_1-r_2) f};y_1|s\biggr)}{ (r_2-x_1) (r_2-x_2) }\right]
\label{131}
\end{eqnarray}
and where we introduced the notation
\begin{eqnarray}
\label{abcd}
    && a=x_1-x_2,\quad b=x_1-x_3,\quad c=x_1-x_4,\quad d=x_2-x_3,\quad f=x_2-x_4,\cr 
    && y_1=  \sin ^{-1}\biggl(\sqrt{\frac{f}{c}}\biggr), \quad y_2=\sin ^{-1}\biggl(\sqrt{\frac{(r-x_1) f}{(r-x_2) c}}\biggr), \quad s=\frac{c d}{b f}.
\end{eqnarray}
One can show that these expressions for $t_q(r)$ and $\phi_q(r)$ satisfy the desired asymptotic behavior, namely, for $r \rightarrow \infty$, we have
\begin{equation}
\phi_q(r) \xrightarrow[\text{$r \rightarrow \infty$}]{} \frac{q}{r} + \mathcal{O}(r^{-3}),
\label{a109}
\end{equation}
and 
\begin{equation}
t_q(r)\xrightarrow[\text{$r \rightarrow \infty$}]{} r +2m \log \left(\frac{r}{2m}\right) + \mathcal{O}(r^{-1}).
\label{a110}
\end{equation}
%

\subsubsection{Parameters of the binary merger}
\label{Sec:Merg}

We are now ready to determine the numerical value of the relevant parameters that characterize the event horizon of the merger of a charged binary black hole system in the EMR limit.
To find the numerical value of the parameters we need to fix the charge to a specific value, which we choose to be $Q=\frac{4}{5}m$. 

\begin{figure}
    \centering
    \includegraphics[width=9cm]{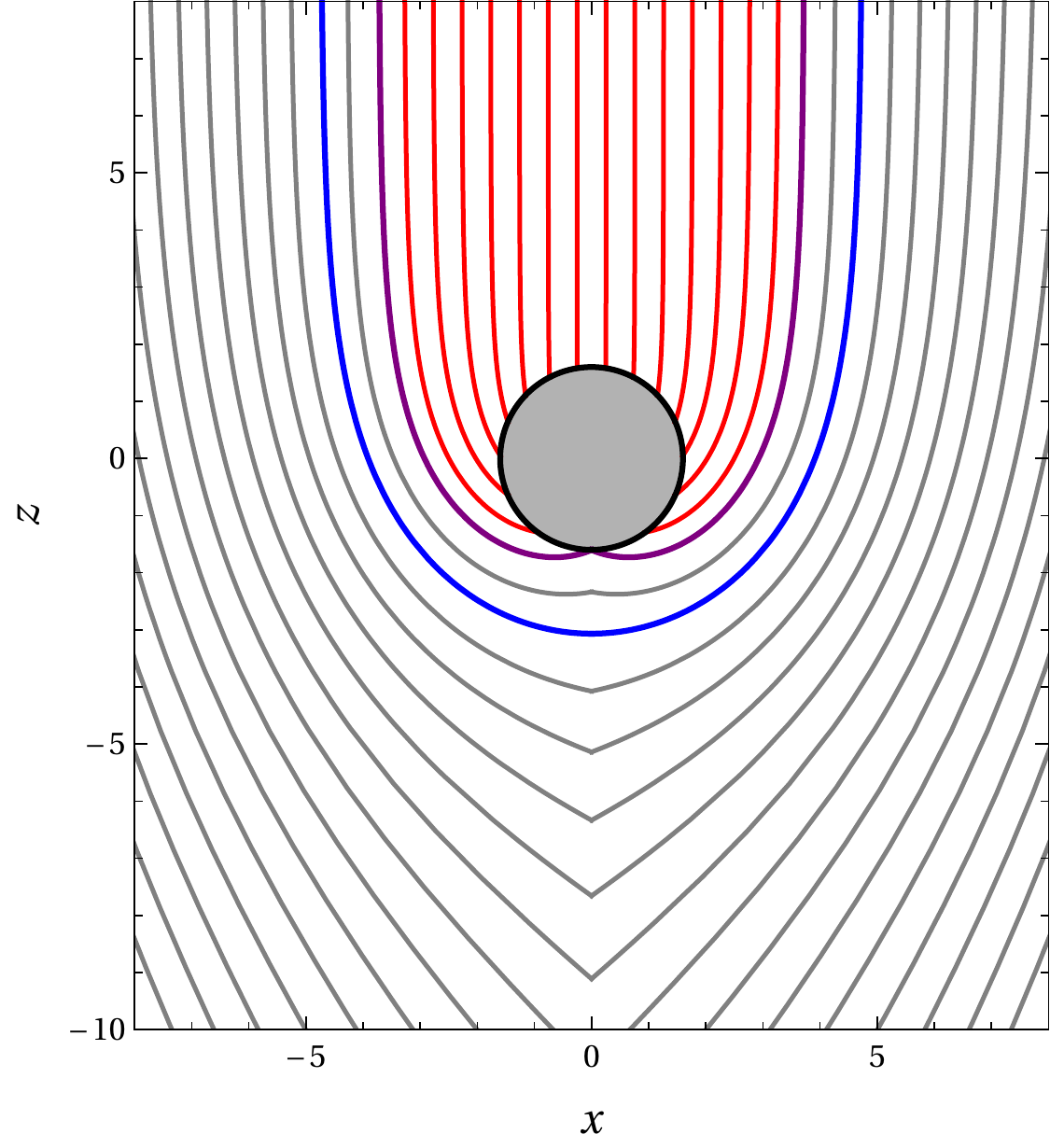}
    \caption{Representation in the $(x,z)$ plane of the event horizon. The grey circle is the RN black hole, the red lines represent the non-caustic generators, the purple ones are the generators with $q=q_c$ and the blue curves are the null geodesics with $q=q_*$. The light rays with $q_c<q<q_*$ enter the event horizon through the small black hole while the generators with $q>q_*$ are able to move away from it. The last generators enter the event horizon through the large black hole. All the curves move towards positive $z$ direction. Note that this representation is analogous to the one obtained for the neutral case studied in~\cite{Emparan:2016ylg} (see Fig. 5 therein).} 
    \label{fig:xz1}
\end{figure}

In order to proceed with our analysis, we distinguish between two classes of generators: the \textit{non-caustic generators} and the \textit{caustic generators}, which are separated by the impact parameter $q=q_c$. The non-caustic generators are the ones characterized by $q \leq q_c$. These are the generators which do not have past endpoints, which means that going back in time, starting from $\mathcal{I}^+$, they will not leave the horizon at a caustic point. On the other hand, caustic generators are the ones with $q>q_c$. These are the ones that enter the horizon at the line of caustics. Among them, we can identify the ones with $q=q_*$, which are the last to enter the event horizon through the caustic line. See Fig.~\ref{fig:xz1} for a graphic representation.

The value $q=q_c$ plays a fundamental role in our computation. It corresponds to the rays that originate from the horizon of the small black hole (i.e. at $r=r_+$) and move in the direction of the large black hole (i.e. $\phi=\pi$). This is similar to the case of the central geodesic, where the light rays start at $r=r_+$ and move in the direction $\phi=0$. The light rays we are interested in here also start at $r=r_+$, but this time they have past endpoints since they do not extend back to infinitely early times. This is possible only if they are moving in the direction of the large black hole, i.e. $\phi=\pi$. Therefore, to determine $q_c$, we need to solve the following equation
\begin{equation}
 \phi_{q_c}(r_+)=\pi.
 \label{132}
 \end{equation}
 We do so numerically and obtain the following value
\begin{equation}
q_c=3.73166m. 
\label{133}
\end{equation}

As stated above, the non-caustic generators are the ones with $q\leq q_c$. We can estimate the growth in the area of the event horizon of the RN black hole considering that its initial area is $\mathcal{A}_{in}=4\pi r_+^2$ and the non-caustic generators contribute at future null infinity with a disk of area $\pi q_c^2$.\footnote{This is simply a consequence of the topology of the three dimensional event horizon at $\mathcal{I}^+$. The null geodesics that define the event horizon and that intersect each other lie on a $S^1$ of radius $q$ at future null infinity.} The growth in the area of this part of the event horizon is 
\begin{equation}
\Delta\mathcal{A}_{\rm{non-caustic}}=\Biggl( \biggl(\frac{q_c}{2r_+}\biggr)^2-1\Biggr)4\pi r_+^2.
\label{134}
\end{equation}
Using this, we obtain 
\begin{equation}
\Delta\mathcal{A}_{\rm{non-caustic}}=0.35989~\mathcal{A}_{in}. 
\label{135}
\end{equation}
It is worth mentioning that in this part of the event horizon no new generators are added. This means that we are considering the growth in the area of the event horizon due to the null geodesics that defined the RN horizon at early times. 

As above, the generators with $q_c<q<\infty$ are called caustic generators. All these generators enter the event horizon at a caustic point at finite time. Among them, the ones with $q=q_*$ are the last to enter the event horizon. The value $q=q_*$, along with $t=t_*$ and $r=r_*$, characterizes the pinch-on instant at which the two horizons merge. To find these values, we follow the light rays with $q=q_*$ back in time from future null infinity. We see that once we arrive at the caustic line they do not approach the small black hole but they also do not leave it. This means that they are ``still" on the collision axis $\phi=\pi$. Thus we have $\dot{r}|_{\phi=\pi}=0$. The null geodesics with $q>q_*$ are instead able to escape the gravitational attraction generated by the small black hole. This means that these light rays will not contribute to the evolution of the event horizon of the RN black hole. On the other hand, the generators with $q<q_*$ cannot escape the gravitational attraction and are forced to move towards the small black hole. With this in mind, it is clear that the last generators to take part in the evolution of the event horizon of the RN black hole are the ones that are unable to escape its gravitational attraction but are not forced to move towards it.

Since in the EMR limit we send $M \rightarrow \infty$, the event horizon of the large black hole becomes infinite. This means that the actual number of null geodesics that define the event horizon at $\mathcal{I}^+$ is infinite. One could then ask what happens to the generators with $q>q_*$, given that the null geodesics with $q=q_*$ are the last to enter the horizon of the small black hole. In this case, it is found that the generators with $q>q_*$ simply enter the final horizon at $\mathcal{I}^+$ through the large black hole.\footnote{Here with final event horizon we mean the event horizon that forms  after the two black holes merge together.}

According to Eq.~\eqref{107}, the values of $r_*$ and $q_*$ can be found by solving 
\begin{equation}
r_*^4-q_*^2r_*^2+2q_*^2mr_*-q_*^2Q^2=0, \hspace{2cm} \phi_{q_*}(r_*)=\pi. 
\label{136}
\end{equation}
Once again we solve these equations numerically and we get %
\begin{equation}
r_*=3.0643m, \hspace{2cm} q_*=4.75396m.
\label{137}
\end{equation}
We can now use these values in the expression for $t_q(r)$ (see Eq.~\eqref{tqr}) to find $t_*$:
\begin{equation}
t_*=-8.10602m. 
\label{139}
\end{equation} 
It is interesting to compare the results \eqref{133}, \eqref{137}, \eqref{139} with the ones of the neutral case analysed in \cite{Emparan:2016ylg}. In the charged case these values are smaller, meaning that the pinch-on happens closer to the centre of the small black hole in the EMR limit. Furthermore, $r_*$ can be taken as a measure of how strongly the small black hole is distorted during the merger. Given that the presence of the charge leads to a smaller $r_*$ than in the neutral case, one can see that a charged black hole is less distorted than a neutral one during the merger.

Another interesting aspect to examine is the duration of the merger. To estimate this, we need to consider the difference $\Delta_*$ between the retarded time at $\mathcal{I}^+$ of the event horizon in the direction $\phi=0$ and the retarded time associated to the light rays emitted when the two black holes merge in the direction of the large black hole ($\phi=\pi$). The latter is simply $t_*$, the former is given by the central generator in Eq.~\eqref{113} evaluated in $r=r_*$ \cite{Emparan:2016ylg}.\\
We define $\Delta_*$ as 
\begin{equation}
\Delta_*=t_{q=0}(r_*) - t_*.
\label{143}
\end{equation}
Inserting in this expression the values of $r_*$ and $t_*$ previously determined, we get
\begin{equation}
\Delta_*=10.4669m.
\label{144}
\end{equation}
The value of $\Delta_*$ in the neutral case is $\Delta_*= 11.89352 m$ \cite{Emparan:2016ylg}. Using the interpretation of $\Delta_*$ as a duration timescale for the merger, it follows that mergers with a charged small black hole are shorter. This comes with no surprise since it was already pointed out in \cite{2020, Liu:2022cuj} that electric and magnetic charges can significantly suppress merger times of charged black hole binaries. This could be explained recalling that a charged black hole is smaller than a neutral one and thus it will be absorbed more quickly by the large black hole.

We can now quantify the change in the area of the small black hole including also the generators with $q_c<q<q_*$. We have
\begin{equation}
\Delta\mathcal{A}_{\rm{smallbh}}=\left(\left(\frac{q_*}{2r_+}\right)^2-1\right)4\pi r_+^2,
\label{145}
\end{equation}
and substituting in the values of $q_*$ we obtain 
\begin{equation}
\Delta\mathcal{A}_{\rm{smallbh}}= 1.20705~\mathcal{A}_{in}.
\label{146}
\end{equation}
Once again, it is interesting to compare the results of the charged case with the ones of the neutral case obtained in \cite{Emparan:2016ylg}. As we can see from \eqref{135} and \eqref{146}, the growth in the area of the event horizon of a charged black hole is much bigger than the growth in the area of the neutral black hole. As we mention in Sec.~\ref{Sec:RN}, the more charge is added, the smaller the black hole becomes. This means that the initial area of a RN black hole of mass $m$ is much smaller than the initial area of a Schwarzschild black hole of the same mass and thus the contribution of the generators (in the growth of the area) is more significant in the charged case. 
Also note that, compared to the case in \eqref{135}, we now need to take into account also the generators with $q_c<q<q_*$. This explains why the results  \eqref{135} and \eqref{146} are different. 
It is also interesting to mention that both \eqref{135} and \eqref{146} show a positive growth in the area of the event horizon, satisfying the second law of black hole mechanics. 

\subsection{Extremal case: $|Q|=m$}
\label{Sec:extr}
The extremal case can be easily studied by taking the limit $Q\to m$ in the expressions \eqref{112}-\eqref{115} or, equivalently, directly in the results given in~\eqref{129}-\eqref{131} derived in the previous section. The expression that we obtain for $t_q(r)$
in this case reads\footnote{Differently from \eqref{tqr} in this expression we already subtracted the integration constant $c_q$.} 
\begin{eqnarray}
 &&t_q(r)=2m \log \left(\frac{q}{2}\frac{\sqrt{(r-x_1) (r-x_2)}+\sqrt{(r-x_3) (r-x_4)}}{\sqrt{(r-x_3) (r-x_4)}-\sqrt{(r-x_1) (r-x_2)}} \right)-x_2+\sqrt{\frac{(r-x_1)(r-x_3)(r-x_4)}{r-x_2}}\cr
 &&+\frac{a m^4}{ \sqrt{b f} (m-x_1)^2 (m-x_2)^2}\biggl[\left(\frac{x_1x_3f-2qm^2+x_2x_4b}{(m-x_4)(m-x_3)}-2(m-q)\right)\Pi \left(\frac{c(m-x_2)}{f(m-x_1)};y_2|s\right)\cr
 &&+\left(2(m-q)(m-x_3)+\frac{bc(m-x_2)^2-df(m-x_1)^2}{2a(m-x_4)}\right)\Pi \left(\frac{c(m-x_2)}{f(m-x_1)};y_1|s\right)\cr
 &&+\frac{b(m-x_1)(m-x_2)}{a(m-x_3)(m-x_4)}\left(f\left(\frac{E(y_1|s)}{2}-E(y_2|s)\right)+\frac{c(m-x_2)}{\sqrt{ab}}\left(4\sin (2y_1)+\frac{\sqrt{(r-x_3)(r-x_2)}}{2(m-r)}\sin (2y_2)\right)\right)\biggr]\cr
 &&+\frac{ma}{\sqrt{bf}}\left(\frac{m^3f}{(m-x_1)(m-x_2)^2(m-x_4)}\left(F(y_2|s)-\frac{F(y_1|s)}{2}\right)+4\left(\Pi \left(\frac{d}{b};y_1|s\right)-\Pi \left(\frac{d}{b};y_2|s\right)\right)\right)\cr
 &&+\frac{\left(\left(x_1x_4+x_2x_3\right) x_2^2-a m \left(4 m^2+(x_1-6 x_2+x_3) m+2 x_2 (x_1+x_3)\right)\right) }{\sqrt{b f} (m-x_2)^2}\left(F\left(y_1|s\right)-F\left(y_2|s\right)\right) \cr
 &&+\frac{m^4}{(m-x_1)^2(m-x_2)^2}\left(\frac{a(2m+q)}{\sqrt{bf}}+\frac{\left(x_1x_3f-2m^2q+x_2x_4b\right)}{2(m-x_3)(m-x_4)}\right)\Pi \left(\frac{c (m-x_2)}{f (m-x_1)};y_1|s\right)\cr
 &&+\frac{m^4}{2(m-x_1)(m-x_4)}\left(\frac{b}{a(m-x_3)}\left(\frac{fE\left(y_1|s\right)}{m-x_2}+\frac{c\sin (2y_1)}{2\sqrt{ab}}\right)-\frac{fF\left(y_1|s\right)}{(m-x_2)^2}\right)
 +\sqrt{b f}\left(E\left(y_1|s\right)-E\left(y_2|s\right)\right)\cr &&
\end{eqnarray}
where we now have that
\begin{equation}
\label{newx}
x_{1,2}=\frac{1}{2}\left(q\mp\sqrt{q(q-4m)}\right),\qquad
    x_{3,4}=\frac{1}{2}\left(-q\mp\sqrt{q(q+4m)}\right),
\end{equation}
and we used the definitions in  \eqref{abcd}.

For the central generator, in the extremal case, we obtain
\begin{equation}
\label{centralextr}
t_{q=0}(r)=r+\frac{Q^2-2m^2}{r-m}+m\log(Q^2+r^2-2mr).
\end{equation}
Note that the expression for $\phi_q(r)$ has the same form as in Eq.~\eqref{129} but where  $x_1, x_2, x_3$ and $x_4$ are instead the ones defined in Eq.~\eqref{newx}.
We can now use the expressions for $t_q(r)$, $t_{q=0}(r)$ and $\phi_q(r)$ to compute the same parameters we extracted in the case $|Q|<m$ discussed in Sec.~\ref{regular}. The geometry of the problem is of course the same therefore, also in the extremal case, we need to solve Eq.~\eqref{132} to determine the value of $q_c$ and Eq.~\eqref{136} to determine $r_*$ and $q_*$. Inserting these last two values in the expression for $t_q(r)$, we get the numerical value of $t_*$, which we can use together with the central generator in Eq.~\eqref{centralextr} (evaluated at $r=r_*$ and for $Q=m$) to extract the duration of the merger $\Delta_*$, according to Eq.~\eqref{143}. \\
The parameters that we obtain in the extremal case are
\begin{equation}
\label{pextr}
q_c=2.69128m,\quad r_*=2.72454m,\quad q_*=4.30440m,\quad t_*=-7.54738m,\quad \Delta_*=9.39568m.
\end{equation}
We can use the values of $q_c$ and $q_*$ to evaluate the growth in the area of the event horizon of the small black hole due to \textit{non-caustic} generators and the \textit{caustic} ones, solving Eq.~\eqref{135} and Eq.~\eqref{145} in the extremal case. We get 
\begin{equation}
\label{areaextr}
    \Delta\mathcal{A}_{\rm{non-caustic}}=0.81075~\mathcal{A}_{in},\quad \Delta\mathcal{A}_{\rm{caustic}}=3.63196~\mathcal{A}_{in}.
\end{equation}
By comparing these results with the ones of Sec.~\ref{regular}, it is clear that in the extremal case the parameters of the merger, i.e. $q_c,q_*,r_*,t_*$, are smaller than the ones obtained in the $Q<m$ case. This means that the small black hole, in the extremal case, is less distorted during the merger~\footnote{This is a consequence of $r_*$ being smaller for an extremal black hole.} and also that the pinch-on happens closer to its center.

From \eqref{pextr} we can also see that the duration of the merger is smaller in this case. This can be explained recalling that an extremal RN black hole is the smallest possible charged black hole, thus it is absorbed quicker by the large black hole during the merger.

Finally we can compare the growth in the area of the event horizon of the small black hole in both cases. From  Eq.~\eqref{areaextr} and Eqs.~\eqref{134} and \eqref{145} it is clear that the contribution of the generators to the area of the event horizon is more significant in the extremal case. Once again this can be explained thanks to the smaller dimension of an extremal black hole compared to a regular RN black hole. 

\subsection{Orbit, spin and charge of the large black hole}
In the previous derivations we have obtained the characteristic equations for the merger without referencing the orbital properties of the binary system nor the charge or spin of the large black hole, which are not necessarily negligible. We will now argue that the results obtained for the radial infall in a non-charged, non-spinning large black hole are valid for any possible merger in the EMR limit. 

Considering first the dynamical parameters, we can immediately see that any relative motion between the black holes in the EMR limit is a linear combination of parallel and perpendicular movement with respect to the event horizon of the large black hole. The invariance under boosts of the asymptotic surface from which the horizon of the large black hole is traced back guarantees the invariance under perpendicular motion; for parallel motion, a similar argument allows us to fix it to zero without loss of generality~\cite{Emparan:2017vyp}. It is also direct to see that the charge of the large black hole is irrelevant in this limit, as the charge term in the RN metric \eqref{RN} scales as $r^{-2}$, whereas the mass term scales as $r^{-1}$. The steeper $r$-dependence of the charge term implies that its contribution becomes irrelevant outside the event horizon of the large black hole in the $M\to \infty$ limit, even if the large black hole was near-extremal.

\section{Numerical solution in $D=4$}
\label{Sec:Num}
\newcommand{\plotwidth}{0.32\textwidth}
\newcommand{\sliceplot}[4]{%
    \ifdraft{}{%
        \begin{figure}[!p]
            \begin{subfigure}{\plotwidth}
                \includegraphics[width=\linewidth]{"AutogeneratedPlots/#2/#3-Plot-9_00"}
                \caption{$t-t_*=-9m$} \label{plot:#1:a}
            \end{subfigure}\hspace*{\fill}
            \begin{subfigure}{\plotwidth}
                \includegraphics[width=\linewidth]{"AutogeneratedPlots/#2/#3-Plot-6_00"}
                \caption{$t-t_*=-6m$} \label{plot:#1:b}
            \end{subfigure}\hspace*{\fill}
            \begin{subfigure}{\plotwidth}
                \includegraphics[width=\linewidth]{"AutogeneratedPlots/#2/#3-Plot-3_00"}
                \caption{$t-t_*=-3m$} \label{plot:#1:c}
            \end{subfigure}
            
            \medskip
            \begin{subfigure}{\plotwidth}
                \includegraphics[width=\linewidth]{"AutogeneratedPlots/#2/#3-Plot+0_00"}
                \caption{$t-t_*=0$} \label{plot:#1:d}
            \end{subfigure}\hspace*{\fill}
            \begin{subfigure}{\plotwidth}
                \includegraphics[width=\linewidth]{"AutogeneratedPlots/#2/#3-Plot+3_00"}
                \caption{$t-t_*=3m$} \label{plot:#1:e}
            \end{subfigure}\hspace*{\fill}
            \begin{subfigure}{\plotwidth}
                \includegraphics[width=\linewidth]{"AutogeneratedPlots/#2/#3-Plot+6_00"}
                \caption{$t-t_*=6m$} \label{plot:#1:f}
            \end{subfigure}
            
            \medskip
            \begin{subfigure}{\plotwidth}
                \includegraphics[width=\linewidth]{"AutogeneratedPlots/#2/#3-Plot+9_00"}
                \caption{$t-t_*=9m$} \label{plot:#1:g}
            \end{subfigure}\hspace*{\fill}
            \begin{subfigure}{\plotwidth}
                \includegraphics[width=\linewidth]{"AutogeneratedPlots/#2/#3-Plot+12_00"}
                \caption{$t-t_*=12m$} \label{plot:#1:h}
            \end{subfigure}\hspace*{\fill}
            \begin{subfigure}{\plotwidth}
                \includegraphics[width=\linewidth]{"AutogeneratedPlots/#2/#3-Plot+15_00"}
                \caption{$t-t_*=15m$} \label{plot:#1:i}
            \end{subfigure}
            
            \caption{#4} \label{plot:#1:sliceplot}
        \end{figure}
    }
}
\newcommand{\xztplot}[5]{%
    \ifdraft{}{%
        \begin{figure}[#5]
            \begin{subfigure}{0.45\textwidth}
                \includegraphics[width=\linewidth]{"AutogeneratedPlots/#2/#3-Plot-xzt1"}
                \caption{} \label{plot:#1:xzt:a} 
            \end{subfigure}\hspace*{\fill}
            \begin{subfigure}{0.45\textwidth}
                \includegraphics[width=\linewidth]{"AutogeneratedPlots/#2/#3-Plot-xzt2"}
                \caption{} \label{plot:#1:xzt:b}
            \end{subfigure}
            \caption{#4} \label{plot:#1:xzt}
        \end{figure}
    }
}

The approach we use for the numerical computation of the event horizon of a charged binary black hole system is slightly different from the analytical case, as we will now work with the Hamiltonian form of the geodesic equations, i.e. using the coordinates $x^\mu$ and the canonical conjugate momenta $p_\mu$. This formalism has the advantage that the resulting differential equations become much easier to solve numerically, mainly because it avoids the problems in the change of signs of the square roots (see \cite{Emparan:2017vyp}). We will use the equations derived in \cite{Emparan:2020uvt}, which we reproduce here for convenience.\footnote{It is important to notice that in \cite{Emparan:2020uvt} the authors work in the perpendicular $\theta=0$ plane and define the geodesics to lie on planes of constant $\phi$. Nevertheless, the spherical symmetry of the problem guarantees that we can always change the non-trivial angular variable $\theta \leftrightarrow \phi$ and get the same form of the geodesic equations.}
\begin{align} 
  \frac{dt}{d\lambda} &= \Delta(r)^{-1} \label{num:1}\\
  \frac{dr}{d\lambda} &= \Delta(r) p_{r} \label{num:2}\\
  \frac{d\phi}{d\lambda} &= \frac{q}{r^2}\label{num:3}\\
  \frac{dp_r}{d\lambda} &= -\frac{\Delta'(r)}{2\Delta(r)^2}-\frac{\Delta'(r)}2 p_r^2 +\frac{q^2}{r^3}\label{num:4}
\end{align}
where, as above, we use $\lambda$ as the affine parameter, $q$ as the impact parameter at infinity, $\Delta(r)=1-\frac{2m}{r}+\frac{Q^2}{r^2}$ and $\Delta'(r)=\frac{d\Delta(r)}{dr}$. 

In order to solve these differential equations, we need a set of integration constants, which we obtain by requiring that the null surface which defines the event horizon at $\mathcal{I}^+$ (i.e. $r\to \infty$) becomes a planar horizon. First, we need to obtain an explicit expression for $p_r(r)$ (i.e. by evaluating $p_\mu p^\mu = 0$)
\begin{equation}
    p_r = \frac{\sqrt{1-\left (1-\frac{2 m}{r}+\frac{Q^2}{r^2}\right)\frac{q^2}{r^2}}}{1-\frac{2 m}{r}+\frac{Q^2}{r^2}} \label{num:5}
\end{equation}
This expression can be used to decouple the inverse equation $d\lambda/dr$. Working with the inverse equation allows us to perform a series expansion around $r\to \infty$, integrate and invert the series. The result yields
\begin{equation}
\label{rlambda}
    r(\lambda)=r_\infty +\lambda +\frac{q^2}{2\lambda}-\frac{m q^2}{2\lambda^2}+\frac{\left (4Q^2-3q^2 \right )q^2}{24\lambda^3}+\mathcal{O}\left (\lambda^{-4}\right)
\end{equation}
Here, $r_\infty$ is the integration constant, which we can set to zero by $\lambda$ reparametrisation. We can then use Eq.~\eqref{rlambda} to solve \eqref{num:2}, \eqref{num:3} and \eqref{num:5} around $\mathcal{I}^+$. We get
\begin{align}
    t(\lambda)&=t_\infty + \lambda + 2 m \log \frac{\lambda}{2 m} + \frac{Q^2-4 m^2 }{\lambda} - \frac{m(8 m^2 - q^2 - 4 Q^2)}{
 2 \lambda^2} \nonumber \\
 &\quad- \frac{16 m^4 + Q^2 (q^2 + Q^2) - 3 m^2 (q^2 + 4 Q^2)}{3 \lambda^3}+\mathcal{O}\left(\lambda^{-4}\right)
\end{align}
\begin{equation}
 \phi(\lambda)=\phi_\infty-\frac{q}{\lambda} + \frac{q^3}{3 \lambda^3}+\mathcal{O}\left(\lambda^{-4}\right)
\end{equation}
\begin{equation}
    p_r(\lambda)=1 + \frac{2 m}{\lambda} + \frac{8m^2 - q^2 - 2Q^2}{2\lambda^2} + \frac{m\left (
 8m^2 - q^2 - 4Q^2\right)}{\lambda^3}+\mathcal{O}\left(\lambda^{-4}\right)
\end{equation}
where the integration constants $t_\infty$ and $\phi_\infty$ can be set to zero without loss of generality by shifting the time origin and orientation of the null plane.

Having determined all the integration constants, we proceed by numerically solving the coupled differential equations \eqref{num:1}, \eqref{num:2}, \eqref{num:3} and 
\eqref{num:4}. From these, we obtain only three independent functions: $t(\lambda)$, $r(\lambda)$ and $\phi(\lambda)$ (as $p_r(\lambda)$ can be computed from $r(\lambda)$ via \eqref{num:5}). Therefore, and since the  parameter $\lambda$ is physically irrelevant, all the non-trivial information about the merger can be summarised in a three-dimensional plot. In order to generate such plots, we use again the cartesian-like coordinates defined in \eqref{38} 
and plot the results in $(x,z,t)$ space. The obtained plots are shown in Fig.~\ref{plot:rn:xzt} and Fig.~\ref{plot:rne:xzt}. It is also worthwhile to generate constant-time slices of these plots, effectively making a movie of the event horizon during the merger. These constant-time slices are shown in Fig. \ref{plot:rn:sliceplot}.

The final step is to prove the agreement between the analytical and numerical methods. In order to do so, we will recompute the merger parameters in Eqs.~\eqref{133}, \eqref{137}, \eqref{139} and \eqref{pextr} using the approach of this section, then check that the values agree. For a regular charged black hole with $Q=\frac{4}{5}m$ (Sec.~\ref{Sec:Merg}), we obtain
\begin{equation}
\label{Numqa}
q_c= 3.73 m, \hspace{2cm}  r_*= 3.07 m, \hspace{2cm} q_*= 4.76 m, \hspace{2cm} t_*= -8.12m, 
\end{equation}
whereas for an extremal black hole (Sec.~\ref{Sec:extr})
\begin{equation}
\label{Numqb}
q_c= 2.71 m, \hspace{2cm} r_*= 2.71 m, \hspace{2cm} q_*= 4.30 m, \hspace{2cm} t_*= -7.57m.
\end{equation}
The two methods agree as expected, with very small discrepancies of the order $\mathcal{O}(0.01m)$.

\xztplot{rn}{RN}{RN}{Event horizon in a merger of a supermassive black hole with a charged black hole of $\left | Q\right |=4/5m$. The red curve represents the caustic line. The axes are measured in units of $m$.}{!p}

\xztplot{rne}{RNExtremal}{RNE}{Event horizon in a merger of a supermassive black hole with a near extremal charged black hole of $m-\left | Q\right |=10^{-13}$. The red curve represents the caustic line. The axes are measured in units of $m$.}{!p}

\sliceplot{rn}{RN}{RN}{Constant-time slices of the event horizon in a merger of a supermassive black hole (down) with a charged black hole of $\left | Q\right |=4/5m$ (centre) in the latter's centre-of-mass reference frame. The event horizon is plotted with a black line, the grey area represents the inside of the black holes. The axes are measured in units of $m$. The time slices are taken at regular time intervals. An animation of these slices can be found in \href{https://github.com/DanielMarinPina/Event-Horizon-of-a-Charged-Black-Hole-Binary-System}{https://github.com/DanielMarinPina/Event-Horizon-of-a-Charged-Black-Hole-Binary-System}} 

\section{Generalisation to $D> 4$ dimensions}
\label{Sec:Dim}
The strength of the numerical computation presented in the previous section is that it can be easily extended to handle more general setups. In this section, we will consider the case of an arbitrary number $D$ of space-time dimensions. Note that this problem is only well defined for $D\geq 4$, as lower-dimensional extensions of general relativity without a cosmological constant do not admit black holes. The study of the event horizon in higher dimensionality is motivated by 
the fact that in string theory and in holographic models, such as the AdS/CFT correspondence (See~\cite{2003hep.th....9246M} and references therein for an introduction to the AdS/CFT correspondence), one typically deals with dimensions higher than 4.
Furthermore, the inclusion of the extra adjustable parameter $D$ allows us to probe which properties of general relativity are inherent to the theory and which are instead artifacts of setting $D=4$.

To generalise our numerical computation, we redefine the function $\Delta(r)$ appearing in the metric as
\begin{equation}\label{num:6}
    \Delta(r)=1-\frac{2 m}{r^{D-3}}+\frac{Q^2}{r^{2(D-3)}}
\end{equation}
For this general case, though, it will not be possible to obtain explicit, $D$-dependent asymptotic conditions. Instead, one would need to perform the asymptotic expansion for each fixed value of $D$. As an example, we will now work in $D=5$ dimensions, for which we obtain the following conditions at $\mathcal{I}^+$
\begin{align}
r(\lambda)&=\lambda +\frac{q^2}{2\lambda}-\frac{mq^2}{2\lambda^2}+\frac{\left(4Q^2-3q^2 \right)q^2}{24\lambda^3}+\mathcal{O}\left (\lambda^{-4}\right)\\
t(\lambda)&=\lambda + \frac{Q^2}{\lambda}-\frac{2 m + Q^2\left(q^2+Q^2\right)}{3\lambda^3}+\mathcal{O}\left(\lambda^{-4}\right)\\
\phi(\lambda)&=-\frac{q}{\lambda} + \frac{q^3}{3 \lambda^3}+\mathcal{O}\left(\lambda^{-4}\right)\\
p_r(\lambda)&=1 +\frac{4 m-q^2}{2\lambda^2}+\mathcal{O}\left(\lambda^{-4}\right)
\end{align}

Similarly as above, we proceed by numerically solving the differential equations \eqref{num:1}, \eqref{num:2}, \eqref{num:3} and \eqref{num:4}, but now making use of the generalised definition of $\Delta(r)$ \eqref{num:6}. As in Sec.~\ref{Sec:Merg}, we choose $Q=\frac{4}{5}m$, but we will now work in $D=5$ dimensions, obtaining the following results, as well as Fig. \ref{plot:rnd:xzt} and \ref{plot:rnd:sliceplot}.
\begin{equation}
q_c^D= 2.50 m, \hspace{2cm}  r_*^D= 1.98 m, \hspace{2cm} q_*^D= 2.72 m, \hspace{2cm} \Delta_*^D=6.26m.
\end{equation}

The most notable difference with the $D=4$ case is that the horizon is much less distorted during the merger ($r_*^D<r_*$) and the merger is shorter ($\Delta_*^D<\Delta_*)$. This is an expected result, as an increase in $D$ implies a steeper $r$-dependence of $\Delta(r)$, so that the characteristic timescale and the length scale for decaying are much shorter. A similar argument explains the relations $q_*^D<q_*$ and $q_c^D<q_c$. In general, one would find that $r_*$, $\Delta_*$, $q_*$ and $q_c$ are monotonously decreasing functions of the number of dimensions, asymptotically approaching zero as $D\to\infty$.

\xztplot{rnd}{RNDim}{RND}{Event horizon in a merger of a supermassive black hole with a charged black hole of $\left | Q\right |=4/5m$ in $D=5$ dimensions. The red curve represents the caustic line. The axes are measured in units of $m$.}{H}

\sliceplot{rnd}{RNDim}{RND}{Constant-time slices of the event horizon in a merger of a supermassive black hole (down) with a charged black hole of $\left | Q\right |=4/5m$ (centre) in the latter's centre-of-mass reference frame and in $D=5$ dimensions. The event horizon is plotted with a black line, the grey area represents the inside of the black holes. The axes are measured in units of $m$. The time slices are taken at regular time intervals. An animation of these slices can be found in \href{https://github.com/DanielMarinPina/Event-Horizon-of-a-Charged-Black-Hole-Binary-System}{https://github.com/DanielMarinPina/Event-Horizon-of-a-Charged-Black-Hole-Binary-System}}

\clearpage
\section{Conclusion}
\label{Sec:conclusions}
In this paper we studied the evolution of the event horizon of a charged black hole binary merger. To do so, we have used the EMR limit, where $m/M \rightarrow 0$. This limit is a very powerful analytical tool which has numerous applications in physics. Usually, it is realised by considering one of the two black holes as a point-like object ($m \rightarrow 0$) while keeping the mass of the other one fixed. In that case, though, the geometrical structure of the small black hole (which defines the event horizon) would vanish, thus making it impossible to study the evolution and the properties of the boundary of the small black hole. Therefore, in the situation we analyse in this paper, the point-particle limit is instead obtained in the complementary case, namely by sending $M \rightarrow \infty$ while keeping $m$ fixed. 

The importance of examining the merger of two black holes in the EMR limit is related to the fact that these mergers do indeed occur in Nature when a small compact object such as a stellar-mass black hole is captured by a super-massive black hole. In the coming years this will be of outmost relevance, since one of the main scientific goals of the upcoming space-based detector LISA is the detection of the gravitational wave signals produced by these mergers \cite{Maselli:2021men, 2015}. Moreover, EMR mergers are important in the investigation of possible extensions of general relativity and the standard model (see e.g. \cite{2019LRR....22....4C} for a status report). 

We performed our analysis of the evolution of the event horizon during the merger both analytically (in Sec. \ref{Sec:RN}) and numerically (in Sec. \ref{Sec:Num} and \ref{Sec:Dim}). The procedure we used, i.e. integrating back in time the null geodesics that define the event horizon at $\mathcal{I}^+$, is completely general and can be used to extract the properties of the event horizon in the merger of any type of black hole. From this, we extracted some of the most relevant features of the merger, such as the presence of a line of caustics, which is general in black hole mergers and allows us to separate the generators of the event horizon in two different classes, caustic and non-caustic generators. We also computed the growth in the area of the event horizon of the small black hole and the duration of the merger. As we can see from Eq.s~\eqref{133}, \eqref{137},  \eqref{139}, \eqref{Numqa} and \eqref{Numqb}, the analytical and numerical approaches agree as expected, with a small discrepancies of order $\mathcal{O}(0.01m)$.

In order to see how the presence of charge changes the parameters of the merger, it is interesting to compare the results of this paper with the ones of Ref.~\cite{Emparan:2016ylg}, which studies the analogous situation for the merger between two neutral black holes in the EMR limit.  In particular, we found that the values of $q_c$, $q_*$, $r_*$ and $t_*$ in our case are smaller than those in the neutral case. The decrease in the radial component can easily be explained as follows: since the event horizon is at $r_+<r_0$ (where $r_0=2m$ indicates the position of the event horizon of a Schwarzschild black hole), the smaller black hole gets less distorted and the pinch-on happens closer to its centre. Moreover, in Sec.~\ref{Sec:RN} we estimated only the growth in the area of the small black hole, neglecting the growth of the horizon of the large black hole. As it is pointed out in \cite{Emparan:2016ylg}, this is a consequence of taking the EMR limit with $M \rightarrow \infty$: in this viewpoint, the size of the event horizon of the large black hole becomes infinite, which means that in order to evaluate the growth in its area we should consider all the infinite number of generators with $q_c<q<\infty$, thing which is highly non-trivial to do in this limit.

The analysis presented in this paper can be extended to include more general scenarios, as for example illustrated in Sec. \ref{Sec:Dim} for the case of an arbitrary number of dimensions. A possible extension of our results to a more realistic and generic situation could consider the case of an EMR binary where the small black hole is a charged rotating black hole described by the Kerr-Newmann solution \cite{Newman:1965my}. In this case, the analysis of the evolution of the event horizon is more complicated due to the spin of the small black hole. A step forward in this direction has been done in \cite{Emparan:2017vyp} for a binary merger in the EMR limit where the small black hole is a neutral, rotating black hole, described by the Kerr metric. The result of this paper could be used as a starting point to study the more difficult general case where the small black hole is described by the Kerr-Newmann metric. As explained in the Introduction, even though black holes are generically considered neutral objects, the inclusion of charge could be relevant in order to investigate how the dynamics of black holes is influenced by its presence and for testing the validity of various exotic models~\cite{Cardoso:2016olt, Preskill:1984gd, Bozzola:2020mjx}.

\section*{Acknowledgments}
We thank Prof. R. Emparan for putting us in contact. We are grateful to Prof. R. Emparan, Prof. G. Grignani and Prof. T. Harmark for useful comments on the manuscript and stimulating discussions. DMP acknowledges financial support by the PRE2020-091801 grant by the State Agency for Research of the Spanish Ministry of Science and Innovation and the FSE+. MO. and DP. acknowledge support from the projects MOSAICO (Fondo Ricerca 2020) and MEGA (Fondo Ricerca 2021) financed by the University of Perugia. MO. and DP. thank the Niels Bohr Institute for hospitality.

\clearpage


\bibliography{Bibliography}
\end{document}